\documentclass[aps,pra,showkeys,twocolumn,superscriptaddress]{revtex4-2}
\usepackage{array}
\usepackage{booktabs}
\usepackage{tabu}
\usepackage{dcolumn}
\usepackage{floatrow}
\usepackage{amsmath}
\usepackage{amsfonts}
\usepackage{float}
\usepackage{amssymb}
\usepackage{graphicx,color}
\usepackage[colorlinks={true}]{hyperref}
\hypersetup{colorlinks=true,linkcolor=red,citecolor=blue,urlcolor=blue}
\usepackage{graphicx}
\usepackage{dcolumn}
\usepackage{bm}
\usepackage{pstricks}
\usepackage{braket}
\usepackage{orcidlink}
\usepackage{lipsum}
\def\be{\begin{equation}}
  \def\ee{\end{equation}}
\def\bea{\begin{eqnarray}}
\def\eea{\end{eqnarray}}
\def\f{\frac}
\def\n{\nonumber}
\def\l{\label}
\def\p{\phi}
\def\o{\over}
\def\R{\rho}
\def\pa{\partial}
\def\om{\omega}
\def\na{\nabla}
\def\P{\Phi}
\begin{document}

\title{Nonequilibrium Quantum Batteries: Amplified Work Extraction Through Thermal Reservoir Modulation} 

\author{Maryam Hadipour \orcidlink{0000-0002-6573-9960}}
\affiliation{Faculty of Physics, Urmia University of Technology, Urmia, Iran}

\author{Soroush Haseli \orcidlink{0000-0003-1031-4815}}\email{soroush.haseli@uut.ac.ir}
\affiliation{Faculty of Physics, Urmia University of Technology, Urmia, Iran}

\date{\today}
\def\be{\begin{equation}}
  \def\ee{\end{equation}}
\def\bea{\begin{eqnarray}}
\def\eea{\end{eqnarray}}
\def\f{\frac}
\def\n{\nonumber}
\def\l{\label}
\def\p{\phi}
\def\o{\over}
\def\R{\rho}
\def\pa{\partial}
\def\om{\omega}
\def\na{\nabla}
\def\P{$\Phi$}

\begin{abstract}
This study examines the steady-state characteristics of work extraction in a two-cell and three-cell quantum battery interacting with multiple thermal reservoirs. Employing the quantum master equation framework within the Born-Markov approximation, we explore the non-equilibrium dynamics governing energy storage and extraction in the system. Our analysis focuses on the influence of thermal gradients across the reservoirs and the impact of inter-cell coupling strength on the battery’s performance. The findings demonstrate that an increase in the middle reservoir temperature substantially enhances the extractable work, underscoring the pivotal role of thermal bath amplification in optimizing energy storage efficiency. Furthermore, we uncover a non-trivial relationship between ergotropy and the coupling strength among the quantum cells, revealing the existence of an optimal coupling regime that maximizes energy extraction. Beyond this threshold, excessive coupling induces energy localization, thereby diminishing the system’s efficiency. These insights provide a theoretical foundation for the strategic design of high-performance quantum batteries by harnessing thermal gradients and interaction-driven control mechanisms.

\end{abstract}

\maketitle

\section{Introduction}\label{intro}
With the rapid advancement of quantum technology in recent years, there has been increasing interest in understanding the thermodynamic principles governing quantum systems\cite{1,2,3,4,5}. This growing focus stems from the need to explore how fundamental thermodynamic concepts, such as energy exchange\citep{6,7}, entropy production \cite{8,9}, and work extraction \cite{10} manifest in the quantum regime. Investigating these aspects is essential for the development of quantum technologies, including quantum heat engines \cite{11,12,13,14} and quantum batteries \citep{15,16,17,18,19,20,21,22,23,24,25,26,27,28,29,30,31,32,33,34,35,36,37,37a}, and for gaining deeper insights into the intricate relationship between quantum mechanics and thermodynamics. 

Given that energy transfer occurs in many quantum processes, quantum thermodynamics provides a theoretical framework for analyzing and quantifying this transfer in quantum systems. A fundamental concept in this field is ergotropy\cite{10}. Ergotropy is defined as the maximum extractable work from a quantum system in a cyclic unitary process.  This concept is particularly significant in assessing the efficiency and limitations of energy conversion in quantum systems, including energy storage and retrieval in quantum batteries. Ergotropy has been invstigated as a valuable resource for performing valuable quantum task \cite{38,39,40}. These investigations have evaluated its role in supplying the necessary energy for executing such processes and analyzed how it can be harnessed within the framework of quantum mechanics. Furthermore, any utilization of ergotropy as an energy source in quantum applications must comply with the energy balance dictated by the first law of thermodynamics and adhere to the fundamental principles of this theory \cite{41}.  
 
However, in an open quantum system interacting with its environment, ergotropy is no longer conserved and is influenced by non-unitary dynamics. The interaction with the environment can lead to energy dissipation, entropy increase, or redistribution of the population among the energy levels of the system, ultimately affecting the amount of extractable work. In open quantum systems, heat and work can be exchanged through thermodynamic processes. This energy transfer, driven by the interaction between the system and its environment, can induce dynamical changes in the system's quantum state. The quantum formulation of the first law of thermodynamic has been introduced by Alicki \cite{41}. In this framework, the internal energy of a quantum system is defined as the expectation value of its Hamiltonian, which governs its dynamical evolution and transformations.

These scientific and theoretical advancements have significantly contributed to diminishing the distinction between quantum thermodynamics and the theory of open quantum systems \cite{42}. In fact, as our understanding of quantum systems interacting with their environment has progressed, it has become evident that the principles of quantum thermodynamics are increasingly intertwined with the concepts of open quantum systems. One of the fundamental topics that has garnered increasing attention in recent research is the optimal design and implementation of both the charging process and the efficient extraction of work from quantum batteries \cite{14,15,16,17,18}. Quantum batteries are quantum systems with a finite energy spectrum that can be controlled and regulated to store energy and transfer it to consumer centers at the appropriate time. In fact, these systems, by utilizing the principles of quantum mechanics, can retain energy in specific states and, when necessary, deliver it to the external environment or a designated receiver. The precise tuning of this process through quantum interactions and the manipulation of system parameters contributes to improving the efficiency of charging and energy extraction in quantum batteries. To date, research has focused on three main approaches for charging quantum batteries: unitary operations \cite{42,43}, coupling to other quantum systems\cite{16,17,18}. In the first approach, the system's time evolution transfers energy to the battery without losing quantum information. In the second, interaction with an external system, such as a environment, allows the system to absorb energy. 
The charging of a quantum battery through thermal mechanisms is a highly intriguing topic, as this approach reduces the need for precise control while enhancing the system's stability \cite{29,45,46}. In this process, thermal energy is transferred from an external source to the battery, and this transfer can occur without the necessity of complex control. Moreover, thermal mechanisms, regarded as a natural means of energy exchange, enable increased efficiency and reduced energy dissipation. It has been shown that the direct thermal charging mechanism of a quantum battery, in which the battery interacts with the environment through the presence of an ancillary system, can lead to the formation of active steady states \cite{47}. When a multipartite system interacts with two or more independent thermal baths, its steady state does not necessarily follow the Gibbs distribution. Under such conditions, non-equilibrium interactions between the composite system and the thermal baths can give rise to active steady states that are far from equilibrium. In general, when a system moves out of thermal equilibrium, heat currents arise between the interacting subsystems. These currents can modify the steady-state population distribution among the eigenstates of the composite system \cite{48} and, under suitable conditions, lead to the emergence of a non-Gibbsian active state \cite{49,50}. \\

In this work, we investigate the steady-state properties and ergotropy of a composite quantum system coupled to multiple thermal reservoirs.  Specifically, we consider a three-qubit system (Quantum battery) with nearest-neighbor interactions, where each qubit is in contact with a thermal bath at a different temperature. The interplay between these baths and the system's internal interactions leads to a nonequilibrium steady state, which we analyze in terms of its energy storage capabilities and thermodynamic properties. We employ the quantum master equation formalism under the Born-Markov approximation to model the dissipative dynamics of the system. The transition rates between system eigenstates are governed by an Ohmic spectral density, and the baths follow a Bose-Einstein thermal occupation distribution. By systematically varying the bath temperatures, we examine the resulting steady-state populations and calculate the ergotropy of the system. Our results provide insights into how nonequilibrium environments can enhance the ergotropy of quantum systems, revealing the potential for optimizing energy storage and work extraction. 
\section{Framework and Nonequilibrium Steady-State Analysis of the System}\label{sec:2}
\begin{figure}[H]
    \centering
  \includegraphics[width = 0.85\linewidth]{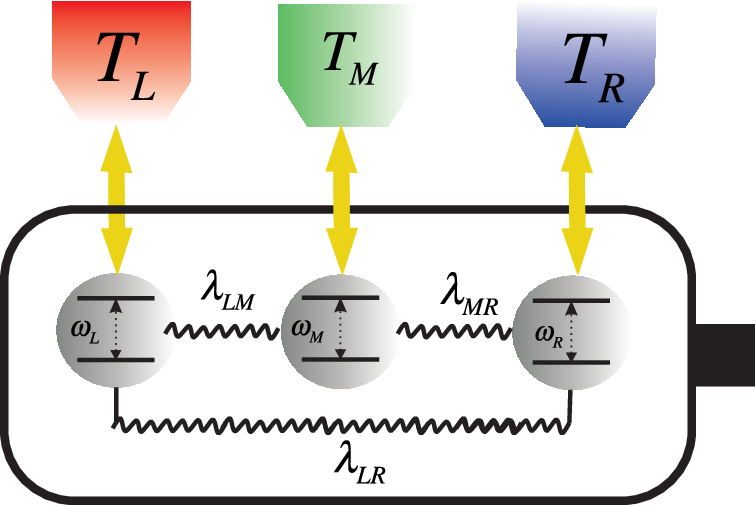}
    \centering
    \caption{The schematic diagram of the quantum battery (QB), composed of three coupled two-level systems interacting with a nonequilibrium environment, featuring left, right, and middle reservoirs.}\label{Fig1}
\end{figure}

The system investigated in this study consists of three two-level systems (TLS) that interact with each other. Additionally, each of these TLS is independently coupled to a thermal bath. This three two-level composite system is considered as (QB). A schematic representation of this model is shown in Fig.\ref{Fig1}. 

In this model, three two-level systems (TLS) are introduced, labeled $L$ (left), $M$ (middle), and $R$ (right). Additionally, each system is coupled to a thermal bath with its corresponding temperature $T_L$, $T_M$ and $T_R$, respectively. The Hamiltonian of the system, which describes the model is 
\begin{equation}\label{h1}
H_S=\sum_{i} \frac{\omega_i}{2} \sigma_z^i + \sum_{i,j \ i\neq j} \frac{\lambda_{ij}}{2} \sigma_z^i \sigma_z^j,
\end{equation}
where $i \in \lbrace L,M,R\rbrace$ and $\sigma_z^i$ corresponds to the $z$-component of the Pauli operator. The eigenstates of the system $i$ associated to $\sigma_z^i$ are the spin up $\vert \uparrow \rangle$ and down $\vert \downarrow \rangle$ states. $\omega_i$ is the transition frequency of the system $i$, while $\lambda_{ij}$ is the coupling strength between system $i$ and $j$. The eigenstates of the system Hamiltonian $H_S$ are constructed as the tensor product of the individual two-level system (TLS) states. Since the system consists of three TLSs, the complete set of eigenstates forms an eight-dimensional basis. These eigenstates are explicitly labeled as 
\begin{equation}
\begin{aligned}
& |1\rangle=|\uparrow \uparrow \uparrow \rangle, \quad| 2\rangle=|\uparrow \uparrow \downarrow\rangle, \quad|3\rangle=|\uparrow \downarrow \uparrow\rangle, \quad|4\rangle=|\uparrow \downarrow \downarrow\rangle, \\
& |5\rangle=|\downarrow \uparrow \uparrow\rangle, \quad|6\rangle=|\downarrow \uparrow \downarrow\rangle, \quad|7\rangle=|\downarrow \downarrow \uparrow\rangle, \quad|8\rangle=| \downarrow \downarrow \downarrow\rangle .
\end{aligned}
\end{equation}
The frequencies associated with the system's allowed energy transitions are given by $\omega_{km}=(\epsilon_k - \epsilon_m)/\hbar$ where each eigenstate $\vert m \rangle$ of the Hamiltonian $H_S$ corresponds to an eigenvalue $\epsilon_m$. The interaction between the two-level system (TLS) and the thermal bath, which consists of a collection of harmonic oscillators, follows the spin-boson model. In this framework, the coupling occurs through the $x$-component of the spin operator. Mathematically, the interaction Hamiltonian can be written as
\begin{equation}
H_I=\sum_{i} \sigma_x^i \otimes \sum_k g_k^i\left(a_k^{i \dagger}+a_k^i\right),
\end{equation}
where $g_k^i$ represents the coupling strength between the $i-th$ TLS  and the 
$k$-th mode of the $i$-th bath, while $a_k^i$ and $a_k^{i \dagger}$ are the annihilation and creation operators of the bosonic bath $i$.  This type of coupling leads to decoherence and energy exchange between the system and the environment, playing a crucial role in quantum dissipation and thermalization processes. The Hamiltonian of the bosonic bath, which consists of a collection of non-interacting harmonic oscillators, is given by $H_B=\sum_{i} \sum_k \omega_k^i a_k^{i \dagger} a_k^i$, where $\omega_k^i$ represents the frequency of the $k$-th mode of $i$-th bosonic bath. This modeling, based on the spin-boson interaction in the $x$-component, implies that the thermal bath can induce transitions between the spin states of the system by flipping one spin at a time. The three TLS undergoes dissipative evolution, meaning that it interacts with its environment in a way that leads to energy exchange and decoherence. By applying the Born-Markov approximation, which assumes weak coupling and no memory effects, the system's density matrix $\rho$ evolves according to a master equation that captures both unitary and dissipative dynamics. The general form of this equation is given by
\begin{equation}\label{mas1}
\dot{\rho}=\mathcal{D}[\rho]=-i\left[H_s, \rho\right]+\mathcal{L}_L[\rho]+\mathcal{L}_M[\rho]+\mathcal{L}_R[\rho] .
\end{equation}
The density matrix $\rho$ satisfies the condition $tr(\rho)=1$, ensuring normalization. The evolution of the system incorporates the Lindblad operator $\mathcal{L}_i(\rho)$, which is given by \cite{51}
\begin{equation}
\mathcal{L}_i[\rho]=\sum_{\omega>0} \mathcal{J}(\omega)\left[\left(1+n_i(\omega)\right) B_i(\omega)+n_i(\omega) C_i(\omega)\right],
\end{equation}
where 
\begin{equation}
\begin{aligned}
& B_i(\omega)=A_i(\omega) \rho A_i^{\dagger}(\omega)-\frac{1}{2}\left\{\rho, A_i^{\dagger}(\omega) A_i(\omega)\right\}, \\
& C_i(\omega)=A_i^{\dagger}(\omega) \rho A_i(\omega)-\frac{1}{2}\left\{\rho, A_i(\omega) A_i^{\dagger}(\omega)\right\},
\end{aligned}
\end{equation}
with $A_i(\omega)=\sum_{\omega>0}|k\rangle\langle k| \sigma_x^i|m\rangle\langle m|$ and $n_i(\omega)=1/(e^{\hbar \omega /\left(k_B T_i\right)}-1)$, for all positive transition frequencies, defined as $\omega_{km}=\omega > 0$ . The thermal baths are considered to be ohmic, meaning that the spectral function takes the form $\mathcal{J}(\omega)=\kappa \omega$, where $\kappa$ is a dimensionless constant \cite{52}. Next, we examine the system in a steady-state regime. As time progresses, the system eventually reaches a steady state, independent of its initial conditions. A dynamic steady state can be obtained by setting the time derivative of $\rho$ to zero in Eq. \ref{mas1}, expressed as $\dot{\rho}=0$. Solving the algebraic equation $\dot{\rho}$ is equivalent to determining the eigenvector of the linear operator $\mathcal{D}[.]$ associated with the zero eigenvalue. We employ the QuTiP library, which provides powerful built-in tools for efficiently solving master equations in the Lindblad form \cite{53,54}. Here, the three TLS is considered as a QB.  The quantum battery undergoes non-unitary charging, but its maximum extractable work is determined through a unitary cyclic process when linked to a consumption hub. The maximum extractable work from a quantum system, whose state is described by the density operator $\rho$ is given by 
\begin{equation}
\mathcal{W}=\operatorname{tr}\left(\rho H_S\right)-\min _U \operatorname{tr}\left(U \rho U^{\dagger} H_{S}\right),
\end{equation}
The optimization is carried out over unitary operators $U$. So, the maximum extractable work from the system is given by
\begin{equation}
\mathcal{W}=tr(\rho H_S)-tr(\pi H_S),
\end{equation}
The above maximum extractable energy is commonly referred to as ergotropy \cite{10}. In above the passive state, denoted as $\pi$, is characterized by the impossibility of extracting work from it. This is expressed by the inequality $tr(\sigma H_S) \leq tr(U \sigma U^{\dagger} H_{S})$. A density matrix  qualifies as passive if it remains diagonal in the Hamiltonian's eigenbasis and its eigenvalues are arranged in a non-increasing order with respect to energy.  This study investigates scenarios in which ergotropy storage resources persist in the steady-state solutions of the system’s dynamics, despite the presence of three thermal baths at different temperatures. In a more general context, the interaction of a composite system with multiple thermal baths can lead to non-equilibrium population distributions and potentially generate coherence within the system's energy eigenbasis. In this study, we demonstrate that a quantum battery can undergo charging when an incoherent heat flow passes through it, arising from the weak coupling of a quantum system to two thermal baths maintained at different temperatures.

In the following, we analyze two distinct scenarios: (i) a quantum battery (QB) composed of two coupled cells, where each cell consists of a two-level system (TLS) individually coupled to a thermal reservoir, and (ii) a QB with three coupled cells, where each cell is also individually coupled to its respective reservoir.
\subsection{Two-cell Quantum battery}
In the first scenario, the quantum battery is considered as a composite system consisting of two coupled two-level subsystems $L$ and $R$. Each subsystem is independently coupled to a thermal reservoir at a fixed temperatures $T_L$ and $T_R$, respectively. This scenario is achieved by setting the Hamiltonian parameters  in Eq. \ref{h1} such that $\omega_L=\omega_R=\omega$, $\lambda_{LM}=\lambda_{MR}=0$ and $\omega_M=0$. So, the Hamiltonian of the system for this scenario can be written as 
\begin{equation}
H_S= \frac{\omega}{2}(\sigma_z^L + \sigma_z^R) + \frac{\lambda_{LR}}{2}\sigma_z^L \sigma_z^R.
\end{equation}
The eigenstates of the system Hamiltonian $H_S$ are $\vert \uparrow \uparrow \rangle$, $\vert \uparrow \downarrow \rangle$, $\vert \downarrow \uparrow \rangle$ and $\vert \downarrow \downarrow \rangle$. So, the master equation describing the dynamics of the system is given by
\begin{equation}\label{mas1}
\dot{\rho}=-i\left[H_s, \rho\right]+\mathcal{L}_L[\rho]+\mathcal{L}_R[\rho] .
\end{equation}
The dynamic steady state can be obtained by setting the time derivative of $\rho$ to zero in above equation. We employ the QuTiP library o solve the equation and obtain  the ergotropy for this scenario. 

\begin{figure}[!h]
    \centering
  \includegraphics[width = 0.85\linewidth]{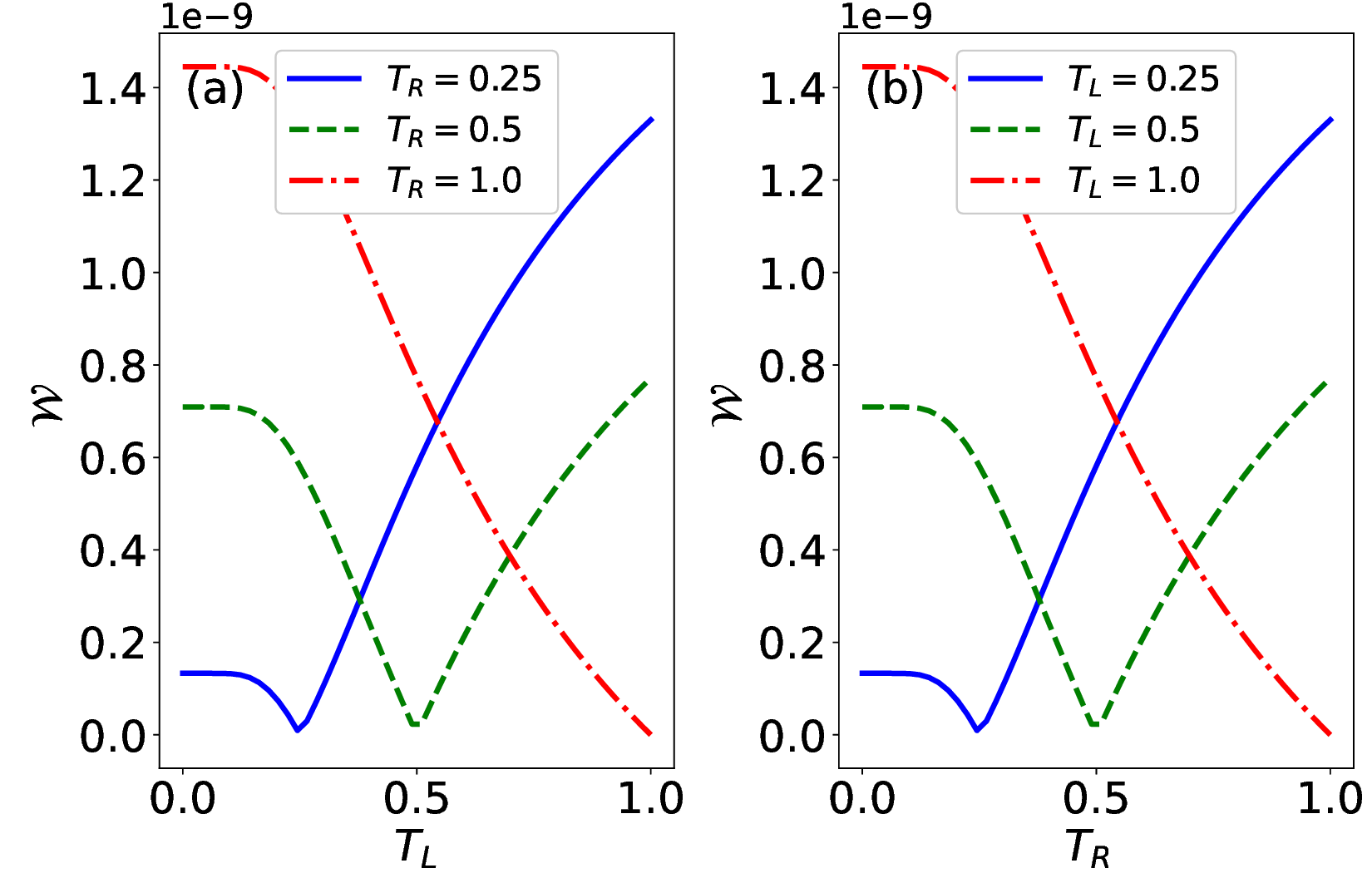}
    \centering
    \caption{(a) The ergotropy $\mathcal{W}$ as a function of the left reservoir temperature $T_L$ for different values of the right reservoir temperature $T_R$. (b) The ergotropy $\mathcal{W}$ as a function of the right reservoir temperature $T_R$ for different values of the left reservoir temperature $T_L$. The other parameters are chosen as: $\lambda_LR$, $\lambda_{LM}=\lambda_{MR}=0$, $\omega_M=0$, $\omega_L=\omega_R=\omega$.}\label{Fig2}
\end{figure}

In Fig. \ref{Fig2}(a), the ergotropy is plotted as a function of the left bath temperature $T_L$ for different values of the right bath temperature $T_R$. In Fig. \ref{Fig2}, the ergotropy is depicted as a function of the temperatures of the left and right baths, denoted as $T_L$ and $T_R$, respectively. From Fig. \ref{Fig2}(a), it can be observed that the ergotropy vanishes in the equilibrium condition $T_L=T_R$.  More precisely, a closer examination of Fig. \ref{Fig2}(a) reveals that the ergotropy is entirely dependent on the temperature difference between the left and right baths $\Delta T = \vert T_L - T_R \vert$. This dependence manifests in such a way that, in the absence of a temperature difference(i.e. $\Delta T=0$), the maximum extractable work from the system is zero. In contrast, as the temperature difference increases, the ergotropy also rises. For example, the solid blue line represents the ergotropy as a function of $T_L$ for $T_R=0.25$. It can be observed that at $T_L=0$, the ergotropy is nonzero. As $T_L$ increases, the ergotropy gradually decreases, reaching its minimum value of zero at $T_L=T_R=0.25$. Beyond this point, as $T_L$ continues to rise and the temperature difference $\Delta T$ increases, the ergotropy starts to grow again. This result indicates that work extraction can be optimized under nonequilibrium conditions. Figure \ref{Fig2}(b) presents the ergotropy as a function of the right reservoir temperature $T_R$ for different values of the left reservoir temperature $T_L$.  The same results are obtained from Fig. \ref{Fig2}(b) as in Fig. \ref{Fig2} (a).

\begin{figure}[!h]
    \centering
  \includegraphics[width = 0.95\linewidth]{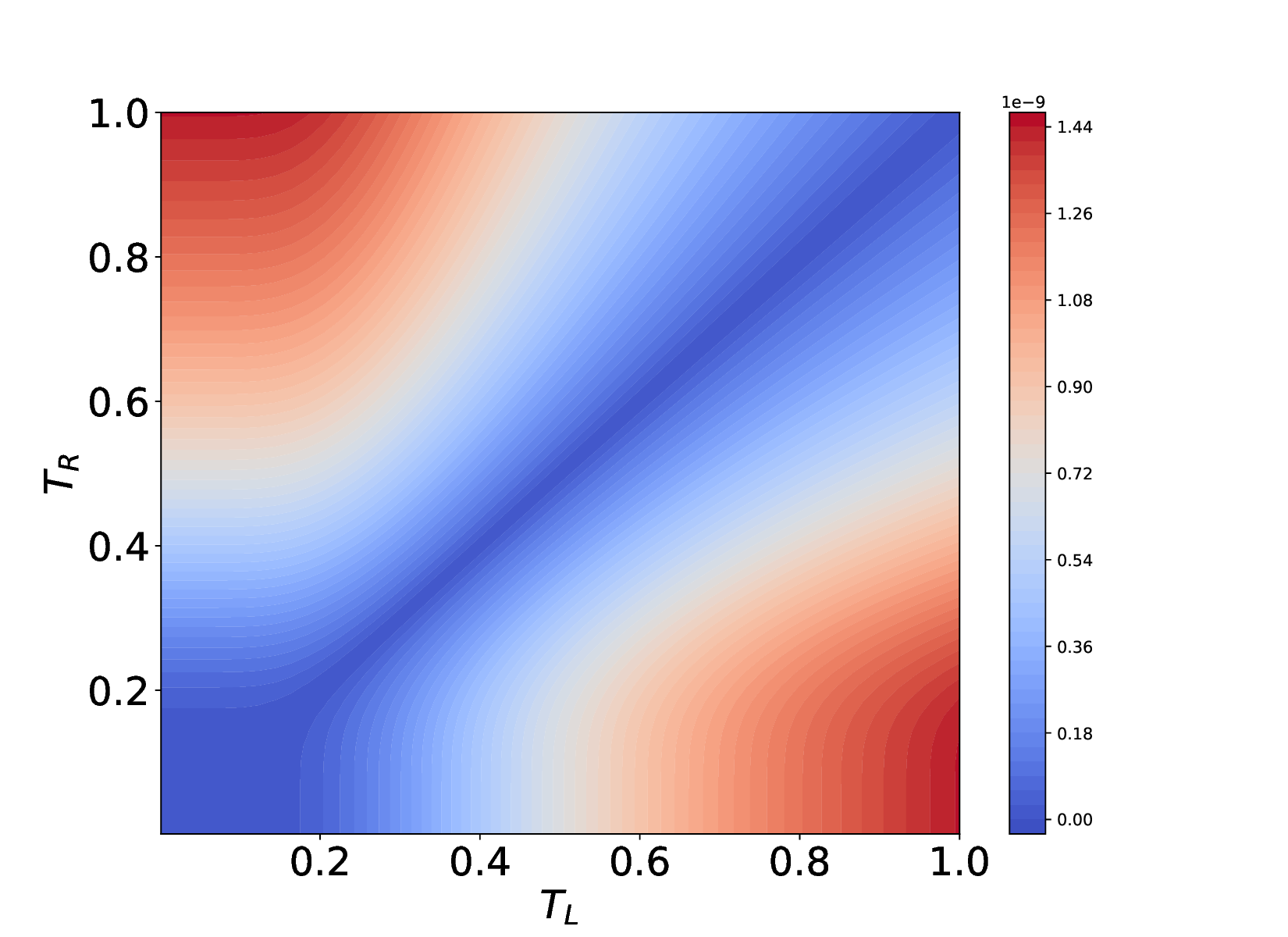}
    \centering
   
    \caption{The ergotropy $\mathcal{W}$ versus left reservoir temperature $T_L$ and right reservoir temperature $T_R$. The other parameters are chosen as: $\lambda_LR$, $\lambda_{LM}=\lambda_{MR}=0$, $\omega_M=0$, $\omega_L=\omega_R=\omega$.}\label{Fig3}
\end{figure} 

To better understand how the ergotropy depends on the temperatures of the left and right reservoirs, a $2D$ plot of ergotropy as a function of both $T_L$ and $T_R$ has been presented in Fig. \ref{Fig3}. This plot also confirms the results obtained from Fig. \ref{Fig2}. As observed in Fig. \ref{Fig3}, the ergotropy reaches its minimum value when the left and right reservoirs are in thermal equilibrium ($T_L=T_R$). However, as the temperature difference $\Delta T$ increases, the amount of extractable work also rises.  So, it can be concluded that the optimal condition for maximizing extractable work, in the case where the system consists of two coupled two-level subsystems interacting with their respective environments, is achieved under nonequilibrium conditions.

\subsection{Three-cell Quantum battery}
In the second scenario, the quantum battery is modeled as a composite system consisting of three coupled two-level systems $L$, $M$ and $R$. Each system is individually coupled to a thermal reservoir at temperatures $T_L$, $T_M$ and $T_R$, respectively.  In this scenario, we consider the general Hamiltonian given in Eq. \ref{h1} and the main dynamical equation of the system in Eq. \ref{mas1} without introducing any further simplifications. The objective of this study is to investigate the impact of the intermediate reservoir's temperature and the presence of an additional cell on the extractable work. To achieve this, we perform a comparative analysis between a two-cell and a three-cell quantum battery. 

\begin{figure}[H]
    \centering
  \includegraphics[width = 0.92\linewidth]{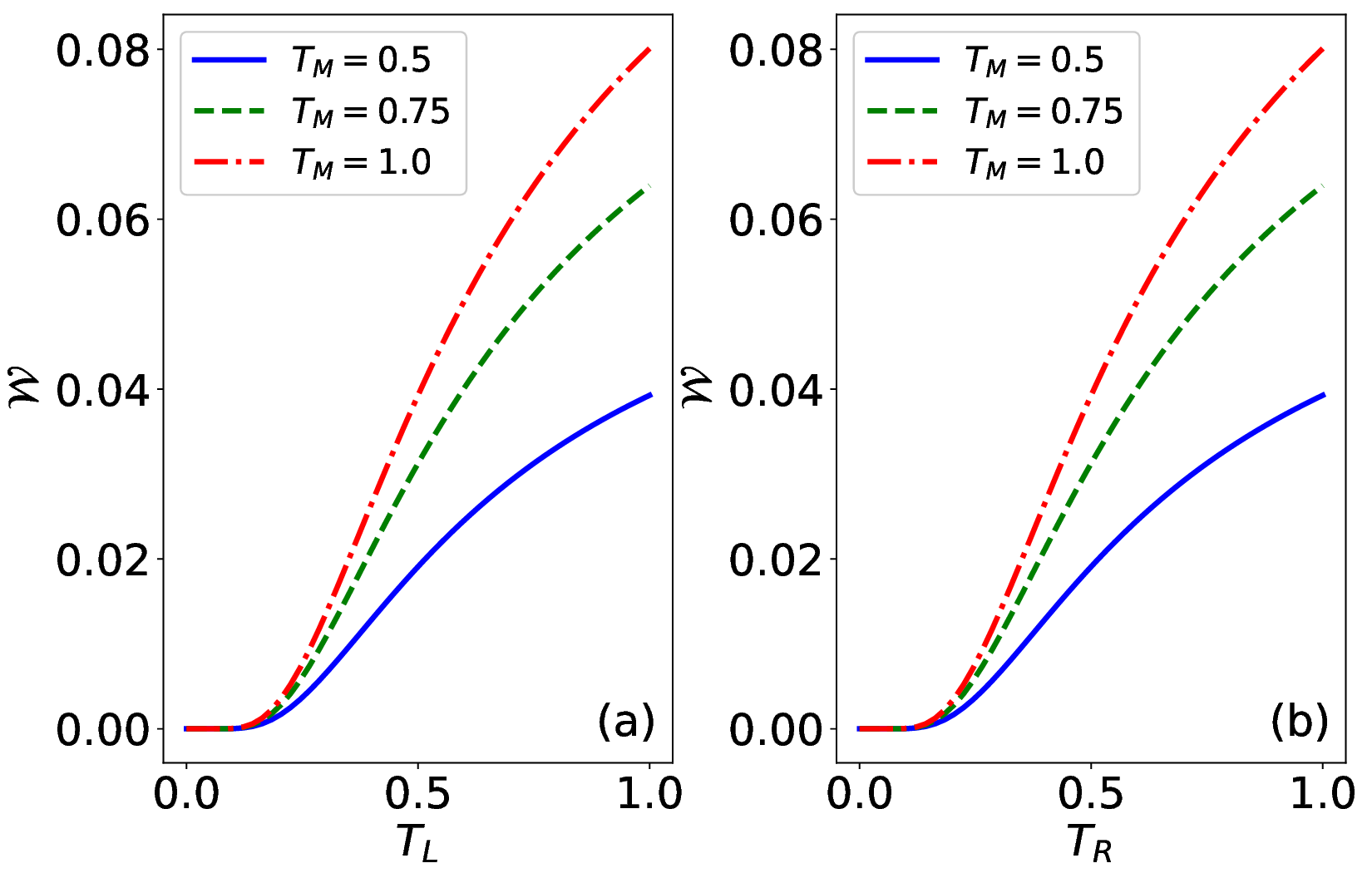}
    \centering
    \caption{(a) The ergotropy $\mathcal{W}$ as a function of the left reservoir temperature $T_L$ for different values of the middle reservoir temperature $T_M$, with $T_R=0$. (b) The ergotropy $\mathcal{W}$ as a function of the right reservoir temperature $T_R$ for different values of the middle reservoir temperature $T_M$, with $T_L=0$, other parameters are $\lambda_{LM}=\lambda_{MR}=\lambda_{LR}=0.1$ and $\omega_{L}=\omega_M=\omega_R = \omega$.}\label{Fig4}
\end{figure}

\begin{figure*}[t]  
    \centering
   \includegraphics[width=\textwidth]{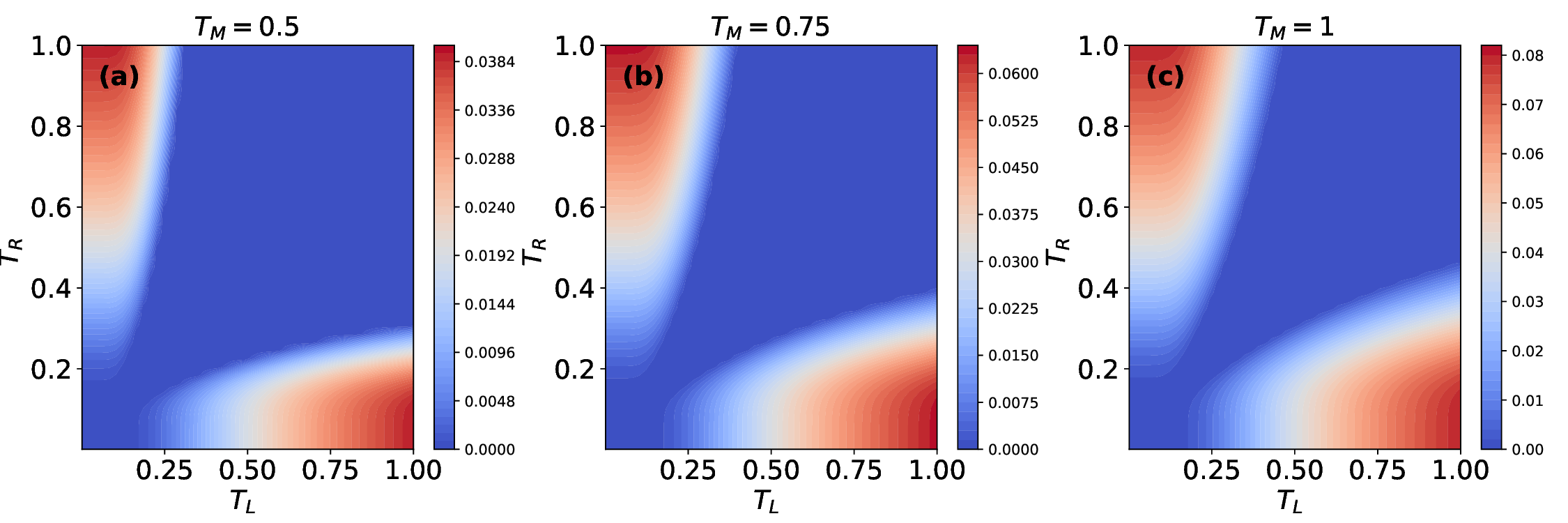}  
    \caption{The ergotropy as a function of the right reservoir temperature $T_R$ and the left reservoir temperature$T_L$ for a three-cell QB. The subfigures correspond to different middle reservoir temperature: (a) $T_M=0.5$ (b) $T_M=0.75$ (c) $T_M=1$. other parameters are $\lambda_{LM}=\lambda_{MR}=\lambda_{LR}=0.1$ and $\omega_{L}=\omega_M=\omega_R = \omega$. }
    \label{Fig5}
\end{figure*}

Fig. \ref{Fig4}(a)  illustrates the variation of ergotropy as a function of the left reservoir temperature  $T_L$ for different values of the middle reservoir temperature $T_M$, while the right reservoir temperature is fixed at $T_R=0$. From Fig. \ref{Fig4}(a), it is observed that as the temperature of the left reservoir rises while the right reservoir temperature remains fixed at $T_R=0$, the thermal gradient between the two reservoirs increases. This greater temperature difference drives stronger energy flows within the system, enhancing the non-equilibrium characteristics that facilitate work extraction. Since ergotropy quantifies the maximum extractable work from a quantum system, an increase in the temperature gradient generally leads to higher ergotropy. This occurs because a larger temperature difference creates a more pronounced imbalance in the population of energy levels, which can be exploited to extract more work from QB. Consequently, as the left reservoir temperature moves further away from that of the right reservoir, the system remains farther from thermal equilibrium, allowing for greater work extraction. It can also be observed that increasing the temperature of the middle reservoir leads to a higher ergotropy. In other words, a higher middle reservoir temperature enhances the performance of the three-cell quantum battery by facilitating greater work extraction. This suggests that the middle reservoir plays a crucial role in boosting the overall efficiency of the system. 

Fig. \ref{Fig4}(b) depicts the ergotropy as a function of the right reservoir temperature for different values of the middle reservoir temperature $T_M$, while the left reservoir is maintained at a fixed temperature of $T_L=0$. From Fig. \ref{Fig4}(b), it is observed that as the temperature of the right reservoir increases and deviates further from the temperature of the left reservoir, the ergotropy increases. This rise is attributed to the growing temperature difference between the left and right reservoirs. Furthermore, the same conclusion drawn from Fig. \ref{Fig4}(a) is confirmed here, namely that increasing the temperature of the middle reservoir enhances the amount of extractable work from the three-cell QB.\\

\begin{figure}[h!]  
    \centering
   \includegraphics[width= 0.85 \textwidth]{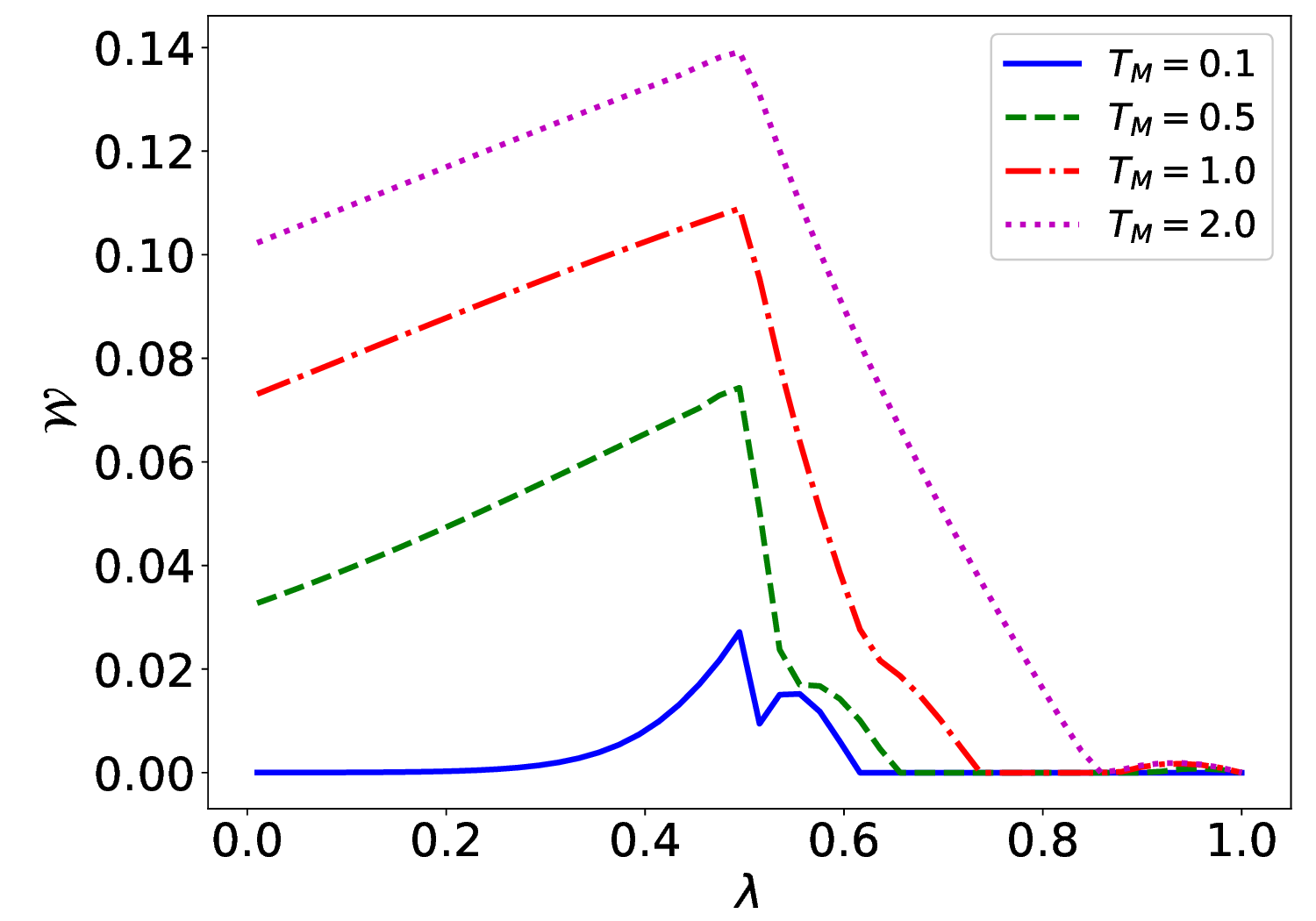}  
    \caption{The ergotropy $\mathcal{W}$ as a function of the coupling parameter $\lambda$ between the battery cells for different values of the middle reservoir temperature $T_M$,other parameters are $T_L=1$, $T_R=0$ and $\omega_{L}=\omega_M=\omega_R = \omega$.  }
    \label{Fig6}
\end{figure}
To gain a deeper insight into the impact of reservoir temperatures on the extractable work of the three-cell quantum battery, a two-dimensional plot is presented in Fig. \ref{Fig5}, depicting the dependence of ergotropy on the right and left reservoir temperatures for three distinct values of the middle reservoir temperature. A comparative analysis of Figs. \ref{Fig5}(a), \ref{Fig5}(b), and \ref{Fig5}(c) reveals that the temperature of the middle reservoir plays a decisive role in enhancing the extractable work from the three-cell quantum battery. The findings indicate that the maximum extractable work is attained when the middle reservoir temperature reaches its highest value.

Another crucial factor that can affect the extractable work from the quantum battery is the coupling strength between the two-level systems. The nature and strength of this coupling play a significant role in determining the efficiency of energy extraction. In this study, we assume that the coupling strengths between all two-level systems are identical, meaning that $\lambda_{LM}=\lambda_{MR}=\lambda_{LR}=\lambda$, ensuring a symmetric interaction among the subsystems. Fig. \ref{Fig6}, illustrates the variation of ergotropy as a function of the coupling parameter between the battery cells for different values of the middle reservoir temperature $T_M$.  At small coupling strengths ($\lambda<<1$), the interaction between the battery cells is weak, meaning that energy remains somewhat localized in individual qubits. As a result, work extraction is limited due to inefficient energy transfer between the cells. As $\lambda$ increases, energy delocalization among the cells improves, allowing for better redistribution of energy and more effective extraction of work. This leads to a gradual increase in ergotropy. At an intermediate value of $\lambda$ (around the value $0.5$), the system reaches an optimal regime where energy is efficiently shared among the quantum battery cells. Beyond a certain threshold of $\lambda$ the ergotropy exhibits a sudden drop. This is due to the formation of strongly correlated states between the battery cells. When the coupling strength becomes  larger the ergotropy decreases and reaches to zero. So, there exist a non-trivial relationship between ergotropy and coupling strength. While moderate coupling enhances energy sharing and maximizes work extraction, excessive coupling leads to energy localization, suppressing ergotropy. This finding underscores the importance of tuning both coupling strength and reservoir temperatures to optimize the performance of quantum batteries.

\section{Conclusion}
In this study, we explored the steady-state properties and ergotropy of a three-cell quantum battery coupled to multiple thermal reservoirs. By employing the quantum master equation formalism within the Born-Markov approximation, we analyzed how temperature gradients and inter-cell coupling influence the system’s energy storage and extraction capabilities.

Our findings reveal that the presence of a middle thermal reservoir significantly enhances the extractable work from the quantum battery. Specifically, increasing the middle reservoir temperature results in a substantial improvement in ergotropy, demonstrating that thermal bath amplification plays a crucial role in optimizing the battery’s performance. This effect highlights the importance of non-equilibrium environments in boosting energy extraction in open quantum systems.

Additionally, we identified a non-trivial dependence of ergotropy on the coupling strength between the quantum battery cells. At weak coupling, energy transfer between cells is limited, leading to inefficient energy extraction. As the coupling strength increases, the system reaches an optimal regime where work extraction is maximized. However, beyond a critical coupling threshold, strong interactions lead to energy localization and a subsequent decline in ergotropy. This behavior is attributed to the formation of strongly correlated states, which restrict the battery’s ability to extract useful work.

Furthermore, our study confirms that the interplay between system-bath interactions and inter-cell coupling plays a fundamental role in determining the steady-state properties of the battery. The results demonstrate that carefully tuning both the reservoir temperatures and the coupling strength can enhance the performance of quantum batteries, paving the way for more efficient quantum energy storage and retrieval mechanisms.

These insights provide a theoretical foundation for the development of high-performance quantum batteries. By leveraging thermal gradients and controlled interactions, future designs can achieve greater efficiency and stability, making them promising candidates for next-generation quantum technologies.

\appendix*

\end{document}